\documentclass[review]{elsarticle}

\usepackage{hyperref}

\journal{Parallel Computing}

\usepackage[utf8]{inputenc}
\usepackage[T1]{fontenc}
\usepackage{lmodern}

\usepackage{textcomp}

\usepackage{algorithm}
\usepackage{algpseudocode}
\makeatletter
\renewcommand{\ALG@name}{Alg.}
\makeatother
\usepackage{amsmath, amsfonts}

\usepackage{booktabs}
\usepackage{rotating}

\usepackage{xcolor}

\usepackage{graphicx}
\graphicspath{{img/}}

\usepackage{nicefrac}

\usepackage{siunitx}
\usepackage{subfig}

\usepackage{tikz}
\usetikzlibrary{decorations}
\usetikzlibrary{decorations.pathmorphing}
\usetikzlibrary{decorations.pathreplacing}
\makeatletter
    \let\pgf@decorate@@brace@brace@code@old\pgf@decorate@@brace@brace@code
    \def\pgf@decorate@@brace@brace@code{
        \ifdim\pgfdecoratedremainingdistance<4\pgfdecorationsegmentamplitude
            \pgftransformxscale{\pgfdecoratedremainingdistance/4\pgfdecorationsegmentamplitude}
            \pgfdecoratedremainingdistance=4\pgfdecorationsegmentamplitude
        \fi
        \pgf@decorate@@brace@brace@code@old
    }
\makeatother
\usetikzlibrary{decorations.shapes}
\usetikzlibrary{shapes}
\usetikzlibrary{shapes.misc}
\tikzset
{
	cross/.style={cross out, draw=black, minimum size=2*(#1-\pgflinewidth), inner sep=0pt, outer sep=0pt},
	cross/.default={1pt} 
}
\newcommand{\particle}[2]
{
	\fill[black!50] (#1,#2) circle [radius = 2];
	\draw (#1,#2) node[cross=3] {};
}
\newcommand{\ghostParticle}[2]
{
	\draw[dashed] (#1,#2) circle [radius = 2];
	\draw[dotted] (#1,#2) node[cross=3] {};
}
\usepackage{url}
\usepackage[colorinlistoftodos]{todonotes}

\newcommand{\walberla}{\mbox{waLBerla}}
\newcommand{\pe}{PE}

\begin{document}

\begin{frontmatter}
\title{A local parallel communication algorithm for polydisperse rigid body dynamics}

\author[erlangen]{Sebastian Eibl\corref{mycorrespondingauthor}}
\cortext[mycorrespondingauthor]{Corresponding author}
\ead{sebastian.eibl@fau.de}
\author[erlangen,cerfacs]{Ulrich Rüde}

\address[erlangen]{Friedrich-Alexander Universit\"{a}t Erlangen-N\"{u}rnberg, Cauerstr.~11, 91058 Erlangen, Germany}
\address[cerfacs]{CERFACS, 42 Avenue Gaspard Coriolis, 31057 Toulouse, Cedex 01, France}

\begin{abstract}
The simulation of large ensembles of particles is usually parallelized by partitioning the domain spatially and using message passing to communicate between the processes handling neighboring subdomains. The particles are represented as individual geometric objects and are assigned to the subdomains. Handling collisions and migrating particles between subdomains, as required for proper parallel execution, requires a complex communication protocol. Typically, the parallelization is restricted to handling only particles that are smaller than a subdomain. In many applications, however, particle sizes may vary drastically with some of them being larger than a subdomain. In this article we propose a new communication and synchronization algorithm that can handle the parallelization without size restrictions on the particles. Despite the additional complexity and extended functionality, the new algorithm introduces only minimal overhead. We demonstrate the scalability of the previous and the new communication algorithms up to almost two million parallel processes and for handling ten billion (\num{e10}) geometrically resolved particles on a state-of-the-art petascale supercomputer. Different scenarios are presented to analyze the performance of the new algorithm and to demonstrate its capability to simulate polydisperse scenarios, where large individual particles can extend across several subdomains. 
\end{abstract}


\end{frontmatter}

\section{Introduction}
\label{sec:introduction}
Many applications in science and engineering require the simulation of a large number of interacting particles. This includes the simulation of macro molecules, powders, mills, silo discharges and more. Due to the insight computer simulations allow into these processes, numerous molecular dynamics frameworks as well as discrete element frameworks like CHARMM~\cite{Brooks1983}, LAMMPS~\cite{Plimpton1995}, GROMACS~\cite{Pronk2013}, NAMD~\cite{Phillips2005},  LIGGGHTS\cite{Kloss2012}, ls1 mardyn~\cite{Horsch2013}, PROJECTCHRONO~\cite{Tasora2016} and waLBerla~\cite{Preclik2015} have been developed. To simulate such a huge number of particles computations may require enormous compute power and memory capacity and can thus be only performed on parallel supercomputers. Here we focus on distributed memory systems that must be programmed with message passing using MPI~\cite{MPI2016}. To fit simulations to this kind of supercomputers the simulation domain is partitioned into subdomains. These subdomains including the associated particles are assigned to different processes which get mapped to the compute cores. Each of these processes is responsible for handling one or more subdomains. For optimal memory use and computational efficiency, each process stores only the information relevant to its own subdomains. This creates the need to exchange information between processes.

In this paper we will focus on rigid particles of finite size which interact via collisions. Note that particles are modeled geometrically, and thus a particle may overlap with two subdomains. In this case, both subdomains need information about this particle and this duplicate information must be kept synchronized. A well known approach is to use nearest-neighbor communication to exchange information between adjacent subdomains. However this approach has clear limitations. When particles are so large that they overlap with several non-neighboring subdomains, communication beyond just the nearest neighbors becomes necessary. The conventional parallelization methods do not support such constellations~\cite{Pinches1991,Liem1991, Rapaport1991, Plimpton1995}. Advanced communication methods improve the communication volume but are also not capable of exchanging information past the next neighbor~\cite{Bowers2005,Bowers2007,Shaw2005,Snir2004,Iglberger2009,Iglberger2009a,Larsson2011,Preclik2015}.

However, this scenario can occur in many practical situations, as e.g. in the simulation of sediments, where the particle size distribution can be assumed to follow a lognormal distribution~\cite{Schruff2016}. Thus a typical simulation will be characterized by many small particles as well as few very large particles. For a fast simulation using a large number of CPUs is desirable and thus the simulation domain will have to be partitioned into many small subdomains. This can easily lead to the case where large particles span multiple subdomains. Such cases cannot be handled correctly by nearest neighbor communication alone, or only at increased cost, when the domain partitioning is modified such that even the largest particles are always contained in the union of only nearest neighbor subdomains. This in turn can lead to an unfavorable load balancing or even to a situation where subdomains become so large that they exceed the memory capacity of a single processor.

A similar problem occurs in coupled simulations when e.g.~suspensions are simulated with the particles embedded in a fluid~\cite{Rettinger2017a,Seil2017}. Then two simulation algorithms, one for particles and one for fluid dynamics, must be co-partitioned and co-scheduled. For efficiency, both methods must use a common domain partitioning. A reasonable load distribution for the flow solver as well as a good load balance for the rigid body dynamics may be impossible to achieve if the size of the largest particle determines a lower bound for the subdomain size.

Finally, also dynamic domain partitioning and refinement algorithms might pose problems. They may induce additional limitations on the domain partitioning as described e.g.\ in \cite{Schornbaum2016} and may lead to equivalent problems as discussed above. In such a constellation with a nonuniform mesh, large moving particles would pose limitations on refinement structure and would limit the flexibility to perform adaptive mesh refinement.

From these examples we conclude that there is significant interest to extend the current nearest-neighbor communication and synchronization algorithm. This article therefore introduces a new {\em shadow-owner} synchronization with the goal to permit more flexible domain partitioning. We describe the new parallel algorithm and analyze its parallel performance systematically. To this end we will show perfect weak scalability as well as agreeable strong scalability. 

This work describes the new synchronization and communication method which has been implemented in the high performance multi physics framework \walberla{} which is openly available, released under GPLv3 license\footnote{\url{walberla.net}}. \walberla{} includes rigid body dynamics as a subsystem called \pe. The excellent scalability and performance of \walberla{}\footnote{\url{www.fz-juelich.de/ias/jsc/EN/Expertise/High-Q-Club/waLBerla/_node.html}} and \pe{}\footnote{\url{www.fz-juelich.de/ias/jsc/EN/Expertise/High-Q-Club/pe/_node.html}} is acknowledged by being a member of the HighQ club of the Forschungszentrum Jülich~\cite{Eibl2017} that includes all programs that have shown scalability to almost half a million compute cores. In this work we demonstrate that the new communication algorithm also scales up to this maximal processor count of a current petascale supercomputer. 

To ensure the comparability of our results the performance data will be expressed in \emph{Particle Updates per Core Second}. This unit is calculated as
\begin{equation*}
	\text{PUpCS} = \frac{\text{total number of simulation steps} * \text{total number of particles}}{\text{total simulation time} * \text{number of cores}}.
\end{equation*}
This enables a comparison between different supercomputer architectures as well as between implementations. The parallel efficiency can also easily be computed from these numbers by normalizing all values to the one for the smallest number of cores.

The remainder of this paper is structured as follows. In section \ref{sec:RigidBodyDynamicsSimulations} we give an introduction to rigid body dynamics simulations detailing the different components needed. We then continue in section \ref{sec:Implementation} with the description of our implementation. We will discuss in detail the common next neighbor synchronization (section \ref{sec:SyncNextNeighbor}) and our newly developed shadow owner synchronization (section \ref{sec:SyncShadowOwner}). In section \ref{sec:PerformanceResults} we discuss the scalability of both algorithms for up to almost 2 million processes as well as the range of scenarios where they can be applied. We then conclude with a summary in section \ref{sec:Conclusion}. 

\section{Rigid Body Dynamics Simulations}
	\label{sec:RigidBodyDynamicsSimulations}
In contrast to molecular dynamics, rigid body dynamics models particles with an actual spatial extent. Interaction between particles takes place when two particles are geometrically in contact with each other. Such collisions must be resolved by suitable models. Since the interaction through collisions is essentially short ranged, many ideas from molecular dynamics for short range potentials can be applied.

The general time step of a parallel rigid body dynamics simulation consists of three parts which are executed in a loop. At first the {\em collision detection} checks which particles are in contact. This is a two step process. First the number of possible contact pairs is reduced. Checking all possible pairs results in a runtime complexity of $\mathcal{O}\left(n^2\right)$ with $n$ being equal to the number of particles which has to be avoided to achieve good performance. Therefore more advanced algorithms like verlet neighbor lists~\cite{Verlet_1967}, cell linked lists~\cite{Hockney1974,Allen2017} or hierarchical hash grids~\cite{Ericson2004,Erleben2005} are used to reduce the complexity to $\mathcal{O}\left(n\right)$. In the second step all candidate pairs identified in the first step are checked for collision using their actual geometry. In the non spherical case an additional selection step can be introduced in between to avoid costly complex collision functions. Commonly axis aligned bounding box (AABB) checks are used for this~\cite{Schinner_1999,Perkins_2001}.

Detected contacts are then passed on to the {\em collision resolution}. Well known algorithms for this stage are soft contact models~\cite{CUNDALL1971,Cundall1979} as well as hard contact models~\cite{Moreau1988,Anitescu1997,Stewart2000}. This stage is also responsible for performing the time integration. For distributed memory parallel simulations an additional {\em synchronization step} is required. Here information must be redistributed such that the data structures are consistent and that every process has all the information it needs to perform the next iteration. The details of this synchronization step depend heavily on the parallelization strategy used. In this paper we focus on the spatial subdivision approach \cite{Plimpton1995,Beazley1995}.

The simulation domain is partitioned into nonoverlapping subdomains and each subdomain is assigned to a processor so that each processor may handle one or several subdomains. This allows the distribution of workload as well as the distribution of data. Each particle is only stored in the subdomain in which its center of mass lies. The location of the particle defines which subdomain and thus which processor {\em owns} the particle. In addition to these \emph{master particles}, we will employ {\em shadow particles} to facilitate distributed memory processing. 

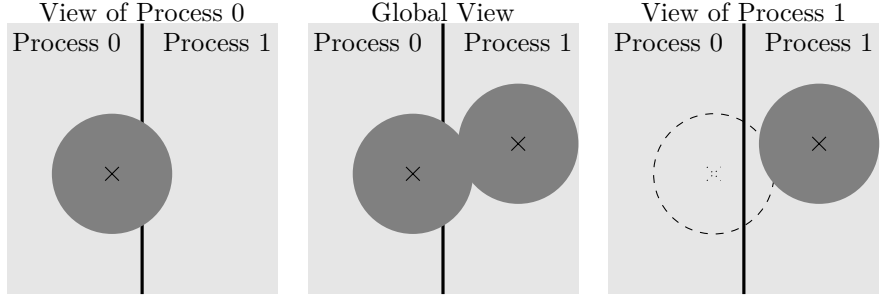
\begin{figure}
		\centering
		\begin{tikzpicture}[scale=0.4]
			\begin{scope}[xshift=-10cm]
				\path [fill = black!10] (-4.5,-5) rectangle (4.5,4);
				\draw (0, 5) node [below] {View of Process 0};
				\draw (-2.5, 4) node [below] {Process 0};
				\draw ( 2.5, 4) node [below] {Process 1};
				\draw [very thick] (0.0,-5.0) -- (0.0,+4.0);
				\particle{-1.0}{-1.0}
			\end{scope}

			\begin{scope}[xshift=0]
				\path [fill = black!10] (-4.5,-5) rectangle (4.5,4);
				\draw (0, 5) node [below] {Global View};
				\draw (-2.5, 4) node [below] {Process 0};
				\draw ( 2.5, 4) node [below] {Process 1};
				\draw [very thick] (0.0,-5.0) -- (0.0,+4.0);
				\particle{-1.0}{-1.0}
				\particle{ 2.5}{ 0.0}
			\end{scope}

			\begin{scope}[xshift=10cm]
				\path [fill = black!10] (-4.5,-5) rectangle (4.5,4);
				\draw (0, 5) node [below] {View of Process 1};
				\draw (-2.5, 4) node [below] {Process 0};
				\draw ( 2.5, 4) node [below] {Process 1};
				\draw [very thick] (0.0,-5.0) -- (0.0,+4.0);
				\ghostParticle{-1.0}{-1.0}
				\particle{ 2.5}{ 0.0}
			\end{scope}
		\end{tikzpicture}
		\caption{
Illustration of the shadow particle concept. Master particles are drawn solid with their centers of mass marked by a cross, whereas the shadow particle is represented by a dotted outline. The situation as seen globally is depicted in the middle. The left and right picture show the information as seen by the two different processes. The shadow particle is necessary so that process 1 can detect the collision that occurs inside its domain with a particle that it does not own.
		}
		\label{fig:SpatialSubdivision}
\end{figure}

The need for shadow particles is illustrated in Fig.~\ref{fig:SpatialSubdivision} showing the situation at the interface between two subdomains that belong to different processes. In order to detect the collision between the two particles one of the two processes must have information about both particles. To this end, the natural choice is to store additional information in each process about particles which overlap the subdomain. This information is organized as \emph{shadow particles} which mirror master particles that are owned by different subdomains. A shadow particle must be created when a particle overlaps a subdomain that is not its owning subdomain. The shadow particle is visualized by a dashed contour in Fig.~\ref{fig:SpatialSubdivision}. Note that using the shadow particle  it is now possible for process 1 to detect the collision and compute a collision response. In the following section we will discuss in detail two alternative synchronization algorithms to manage the shadow particles.

\section{Parallelization using the \walberla{} Framework}
	\label{sec:Implementation}
waLBerla is a software framework that provides functionality for domain partitioning using cuboidal subdomains that are organized in a forest of octrees \cite{Godenschwager2013, Schornbaum2016}. Excellent scalability and efficiency is achieved since all data structures are stored in a fully distributed manner avoiding storing any global information. In particular each process has only information about its own subdomains and its directly adjacent neighbors. This includes particle data and domain partitioning information.

\subsection{General Requirements for Synchronization Algorithms}
\label{sec:Requirements}
The two synchronization methods discussed in this paper are independent of the contact model. As long as the interaction is short ranged they can be used for both molecular dynamics as well as rigid body dynamics. Short ranged in this context means that particles only interact with each other when they are closer than a certain threshold. This threshold has to be much smaller then the domain size. In this paper we focus on rigid body dynamics using a non smooth granular dynamics model~\cite{Preclik2015}. For both algorithms we rely on the assumption that particles do not move farther than half the diameter of the smallest particle within one time step. Since the physical accuracy of the simulation would decrease for even lower velocities this is no constraint.

When particles are generated, the respective memory is allocated in the subdomain where their center of mass lies. The synchronization algorithm must then create and update shadow particles whenever they are needed on other processes. Shadow particles hold {\em passive} information that can be read by the algorithms, but not modified. In particular, all geometric transformations, such as translations and rotations that occur due to time integration are executed only on the master particle and are subsequently synchronized to all shadow particles. The synchronization is also responsible for deleting shadow particles when they are not necessary anymore. For all these tasks, the synchronization algorithm maintains a list of shadow particles for every master particle on the process of the master particle. Additionally a reference back from every shadow particle to its corresponding master particle is stored on the process of every shadow particle. This data structure enables to efficiently update shadow particles and to collect and aggregate interactions. Since it is sometimes unavoidable to set forces and torques on shadow particles -- either due to user interaction or due to collisions -- these contributions are cached and collected on the master particle during the collision resolution phase. The reduction algorithm used for this can also rely on the references maintained by the synchronization algorithm.

The different types of information are communicated with a special messaging protocol. Each message consists of an identifier and message specific data. Both the identifier as well as the data are packed at the sender side and are unpacked to be processed at the receiver side. To optimize communication time, messages sent to the same destination process are first collected and sent in one block. This message aggregation helps to reduce the impact of latency involved in every communication. A more detailed description of this optimized message protocol can be found in~\cite{Iglberger2009}.

In summary, synchronization algorithms must fulfill three tasks:
	\begin{itemize}
		\item create and delete shadow particles
		\item update shadow particles when master particle is modified
		\item maintain references between corresponding master and shadow particles
	\end{itemize}
Two additional requirements must be met to achieve good scaling results:
	\begin{itemize}
		\item use only local and next neighbor information
		\item reduce non next neighbor communication to a minimum
	\end{itemize}
We will refer to these requirements as \emph{simulation requirements} and \emph{scaling requirements}. In the following two sections a basic synchronization algorithm as well as our extension will be discussed.

\subsection{Next Neighbor Synchronization}
\label{sec:SyncNextNeighbor}
\begin{algorithm}
	\caption{Next Neighbor Synchronization}\label{alg:NNS}
	\begin{algorithmic}[1]
		\Function{syncNextNeighbor}{}
			\ForAll{master particles}
				\If {overlaps neighboring domain} \Comment{step 1}
					\If {neighbor is registered as shadow owner}
						\State \Call{update properties of the corresponding shadow particle}{}
					\Else
						\State \Call{create new shadow particle}{}
						\State \Call{register new shadow owner}{}
					\EndIf
				\Else
					\If {neighbor is registered as shadow owner} \label{alg:NN:reg}
						\State \Call{remove shadow particle}{}
						\State \Call{deregister shadow owner}{}
					\EndIf
				\EndIf
				\If {particle left own subdomain} \Comment{step 2}
					\State \Call{find new owner}{}
					\If {new owner found}
						\State \Call{transfer ownership}{}
						\State \Call{update owner of all shadow particles}{}
					\Else
						\State \Call{remove particle and all its shadow particles}{}
					\EndIf
				\EndIf
			\EndFor
			\State \Call{send\&receive messages}{} \Comment{all messages are aggregated till this point}
			\EndFunction
	\end{algorithmic}
\end{algorithm}

For the basic next neighbor synchronization (NNS) one assumption has to be made. The radius of the largest particle, i.e. the maximal distance between the center of mass and the confining hull, has to be smaller than the smallest edge length of the subdomains. If this is true one can achieve all three tasks within one communication step by only using next neighbor communication. This communication pattern is for example described in~\cite{Rapaport1991}.It is very well suited for highly parallel computations and exhibits perfect weak and strong scaling~\cite{Eibl2017}.

We will now describe the algorithm in detail. A pseudocode version is shown in Algorithm~\ref{alg:NNS}. Since only next neighbors are involved every aspect can be controlled by the owner. Therefore every subdomain processes only its master particles. Everything related to shadow particles must be communicated to the corresponding process. This involves sending a message. The algorithm can be divided into two steps. In the first step, for all master particles that overlap a neighboring subdomain which does not have a shadow particle, one is created. If there is already a shadow particle all its properties get updated with the values from the master particle. These two cases are treated separately since updating involves less data to be sent in comparison to a full copy. If there is no overlap but a shadow particle exists then it gets deleted. In all these operations the references between the master and shadow particles are updated to remain consistent.

In the second step for all master particles it is checked whether their center of mass still lies within the owning subdomain. If it has left its subdomain, the new owner is determined. The new owner will subsequently promote its shadow particle to a master particle and all other shadow copies get notified about the new owner. If no owner is found - this means that the particle has left the simulation domain - it gets deleted. Since there is no dependency between the messages within this algorithm, they can be aggregated and sent in a block at the end of the routine.

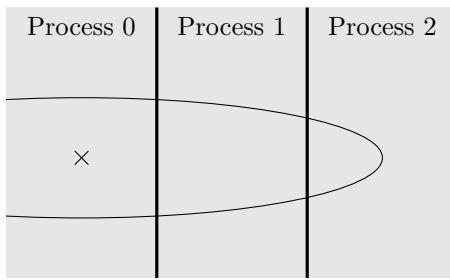
\begin{figure}
		\centering
		\begin{tikzpicture}[scale=0.4]
			\tikzset{cross/.style={cross out, draw=black, minimum size=2*(#1-\pgflinewidth), inner sep=0pt, outer sep=0pt},
cross/.default={1pt}}

				\path [fill = black!10] (-7.5,-5) rectangle (7.5,4);
				\draw (-5, 4) node [below] {Process 0};
				\draw ( 0, 4) node [below] {Process 1};
				\draw (+5, 4) node [below] {Process 2};
				\draw [very thick] (-2.5,-5.0) -- (-2.5, +4.0);
				\draw [very thick] ( 2.5,-5.0) -- (2.5,+4.0);
				\clip (-7.5,-5) rectangle (7.5,4);
				\draw (-5,-1) circle [x radius = 10, y radius = 2];
				\draw (-5,-1) node[cross=3] {};
		\end{tikzpicture}
		\caption{The picture illustrates the case where one particle is large enough to reach non adjacent subdomains. In this case the next neighbor synchronization cannot be used since Alg.~\ref{alg:NNS} contains no communication between non adjacent subdomains.}
		\label{fig:NNfail}
	\end{figure}

A significant drawback of this algorithm is that if the initial assumption about the maximal particle size does not hold, the synchronization method cannot be used (see Fig.~\ref{fig:NNfail}). Since it only involves next neighbor communication subdomains farther away will not be updated. This would eventually result in undetected collisions and wrong simulations. A correct simulation would require sufficiently large subdomains. However the subdomain size might be restricted by coupled algorithms or load balancing algorithms like described in the introduction. An alternate synchronization method is needed to tackle that case. In the next section we describe our new approach which does not depend on the next neighbor assumption.

\subsection{Shadow Owner Synchronization}
	\label{sec:SyncShadowOwner}
		\begin{algorithm}
			\caption{Shadow Owner Synchronization}\label{alg:SOS}
		\begin{algorithmic}[1]
			\Function{syncShadowOwners}{}
			\ForAll{master particles}
				\State \Call{update properties of all shadow particles}{}
				\If {particle left own subdomain}
					\State \Call{find new owner}{}
					\If {new owner found}
						\State \Call{transfer ownership}{}
						\State \Call{update owner of all shadow particles}{}
					\Else
						\State \Call{remove particle and all its shadow particles}{}
					\EndIf
				\EndIf
			\EndFor
			\State \Call{send\&receive messages}{} \Comment{all messages are aggregated till this point}
			\ForAll{particles}
				\If {overlaps neighboring domain}
					\If {neighbor is not owner and not cached}
						\State \Call{create new shadow particle}{}
						\State \Call{register new shadow owner}{}
						\State \Call{flag neighbor}{}
					\EndIf
				\EndIf
				\If {particle is a shadow particle and no longer touching the current subdomain}
					\State \Call{delete particle}{}
					\State \Call{notify neighbors and owner}{}
				\EndIf
			\EndFor
			\State \Call{send\&receive messages}{} \Comment{all messages are aggregated till this point}
			\EndFunction
		\end{algorithmic}
		\end{algorithm}

When particles do not satisfy the next neighbor assumption, one option is to handle them globally. This means that every process holds a copy of these large particles and all synchronizations involve global communication. Since global communication does not scale with the number of processes involved, this would degrade the parallel efficiency in large scale simulations. Therefore our alternative approach avoids global communication entirely and instead uses an approach which we call diffusive. The idea behind this is to propagate information to neighbors if they are lacking it. This way information spreads like a drop of ink in water, slowly covering the whole domain. Applied to our synchronization problem, every subdomain checks its neighbors if they need a shadow particle for one of its master or shadow particles. If so they send the required information. This check can be executed locally and does not involve global communication. Information about shadow particles will propagate one subdomain per synchronization call. Also the update of shadow particles can be performed by point-to-point communication using the list of shadow particles maintained for the master particle. Though more expensive than simple next neighbor communication this avoids an expensive global communication.

Note that information is propagated by each subdomain based only on its local view of its neighborhood. Therefore the situation may arise that one subdomain receives the same information twice from different senders. However, in all cases one is the exact duplicate of the other and one is discarded.

When a large particle that covers more than one subdomain is initialized, more synchronization calls are required to create all shadow particles. In every synchronization call new shadow particles can be created in neighboring subdomains of existing shadow particles. Therefore the exact number of synchronizations calls is dependent on the size of the particle. Once all initial references between master and shadow particles have been created one synchronization call per time step is sufficient to keep all references up to date. In the following we will refer to this new algorithm as {\emph{shadow owner synchronization (SOS)}}.

A pseudo code with the details of the SOS can be found in Algorithm~\ref{alg:SOS}. The algorithm can be decomposed into two parts. The first part starts with updating all shadow particles with values from the master particle. Then the ownership of all master particles is checked by finding the subdomain which contains the center of mass. When needed the ownership is transfered. The second part of the algorithm checks all particles - master as well as shadow particles - if they overlap an adjacent subdomain. If needed and not already existent new shadow particles are created. The creation also involves an update of the shadow owner list stored by the master particle. Then every subdomain deletes shadow particles that are not needed anymore.

In this algorithm synchronization is needed at two points. All messages in the first part can be aggregated. The communication pattern is a point-to-point communication between the master particle and all its shadow particles. The second point of communication is after the second part of the algorithm. Again all messages can be collected and sent as one large aggregated block. This communication is again a point-to-point communication between master and shadow particles with next neighbors added. In total this algorithm needs two communication steps and the number of messages sent is roughly proportional to the number of shadow particles.{}

In its basic form this algorithm will produce unnecessary communications. Since shadow copies do not have any information about the other shadow copies their subdomains will attempt to generate new shadow copies at adjacent subdomains as part of every synchronization. To avoid this overhead, a cache is introduced for every subdomain. This cache stores information about previously executed communication. In particular, if the creation of a shadow particle on a neighboring subdomain has been requested. With this information the cache can be used to avoid redundant communication (compare line 17 of Alg. \ref{alg:SOS}). The cost of this optimization is the additional memory needed as well as management overhead. However, since this cache only stores information about neighboring subdomains its size and cost remains bounded and thus supports scaling to large processor numbers. For further details see the pseudo code in Alg.~\ref{alg:SOS}.

The proposed synchronization algorithm fulfills all simulation as well as scaling requirements stated in Sec.~\ref{sec:Requirements}. Scaling results for this algorithm are discussed in the next section.

\section{Performance Results}
\label{sec:PerformanceResults}
In this section the scalability of both synchronization algorithms is evaluated and the advantages as well as disadvantages of our approach are discussed. All performance measurements are conducted on the Juqueen supercomputer located at the Jülich Supercomputing Centre (JSC)~\cite{juqueen}. This is a IBM BlueGene/Q cluster with \num{28672} nodes. Each node contains one PowerPC A2 processor with \num{18} (only 16 are usable) cores clocked at \SI{1.6}{GHz}~\cite{Gilge2014,Wautelet2014}. The processors support 4-way simultaneous multithreading (SMT) for all cores. Previous tests showed that making full use of the multithreading capabilities yields the best performance~\cite{Eibl2017}. For full-machine jobs this sums up to \num{1835008} processes. Each node also offers \SI{16}{GB} of RAM. The nodes are connected via a 5D torus network.

\subsection{Weak and Strong Scalability}
To demonstrate the scalability of these algorithms we perform weak- and strong-scaling experiments. In weak-scaling experiments the number of processes is increased together with the problem size. For perfect weak-scaling we expect the time to solution to stay constant. The performance unit PUpCS will reflect that by also staying constant since the increase in the number of particles will cancel out with the increase in the number of cores. Let $t_p$ be the time to solution for p processes. For weak-scaling experiments the parallel efficiency is then defined to be
\begin{equation*}
e_{p,ws} = \frac{t_1}{t_p}.
\end{equation*}

For strong-scaling experiments the number of cores is increased but the problem size stays the same. In this case, ideally, the computation time should decrease by the same amount as the computation power is increased. Again this will have no effect on the PUpCS measure since both contributions will cancel out each other. With the same definition of $t_p$ as before the parallel efficiency for strong-scaling is defined to be:
\begin{equation*}
e_{p,ss} = \frac{t_1}{pt_p}.
\end{equation*}

More in-depth information about performance measurements in general can be found in~\cite{Hager2010}.

\subsection{Scalability in Different Scenarios}
	\label{sec:Scaling}{}
	\begin{figure}
		\centering

		\subfloat[sparse scenario]{
		\begin{tikzpicture}[scale = 0.4]
		\foreach \j in {0,...,4}
		{%
		    \foreach \i in {0,...,4}{\fill[black!50] (\i, \j) circle [radius=0.4];}   
		}
		\draw[very thick, dashed] (-0.5, -1) -- (-0.5, 5);
		\draw[very thick, dashed] (+4.5, -1) -- (+4.5, 5);
		\draw[very thick, dashed] (-1, -0.5) -- (+5, -0.5);
		\draw[very thick, dashed] (-1, +4.5) -- (+5, +4.5);
		\end{tikzpicture}
		}
		\hspace{1cm}
		\subfloat[dense scenario]{
		\begin{tikzpicture}[scale = 0.4, rotate = 60]
		[hexa/.style= {shape=regular polygon, regular polygon sides=6, minimum size=1cm, draw,anchor=south}]

		\foreach \j in {0,...,3}{%
		    \ifodd\j 
		         \foreach \i in {0,...,4}{\fill[black!50] ({\j * sin(60)},{(\i+0.5)}) circle [radius=0.5];}   
		    \else
		         \foreach \i in {0,...,4}{\fill[black!50] ({\j * sin(60)},{\i}) circle [radius=0.5];}
		    \fi}
		\draw[very thick] ({-0.0 * sin(60) - 0.5}, -1) -- ({-0.0 * sin(60) - 0.5}, 5);
		\draw[very thick] ({+3.0 * sin(60) + 0.5}, -1) -- ({+3.0 * sin(60) + 0.5}, 5);
		\draw[very thick, dashed] ({-1 * sin(60)}, 0) -- ({+4 * sin(60)},0);
		\draw[very thick, dashed] ({-1 * sin(60)}, 4) -- ({+4 * sin(60)},4);
		\end{tikzpicture}
		}
		\hspace{1cm}
		\subfloat[large scenario]{
		\begin{tikzpicture}[scale = 0.4]

		\foreach \j in {0,...,4}
		{
		    \fill[black!50] (0, \j) circle [radius=0.4];  
		    \fill[black!50] (4, \j) circle [radius=0.4];  
		}
		\foreach \j in {0,...,4}
		{
		    \fill[black!50] (\j, 0) circle [radius=0.4];  
		    \fill[black!50] (\j, 4) circle [radius=0.4];  
		}
		\fill[black!50] (2,2) circle [radius=1.5];
		\draw[very thick, dashed] (-0.5, -1) -- (-0.5, 5);
		\draw[very thick, dashed] (+4.5, -1) -- (+4.5, 5);
		\draw[very thick, dashed] (-1, -0.5) -- (+5, -0.5);
		\draw[very thick, dashed] (-1, +4.5) -- (+5, +4.5);
		\end{tikzpicture}
		}

		\caption{Illustration of a 2d equivalent of the sparse a), dense b) and large c) scenario. These illustrations contain far less particles than the real scenarios and the scaling is not correct. Each illustration depicts a building block which is repeated periodically at its boundaries. Periodic boundaries are marked as dashed lines.}
		\label{fig:Scenarios}
	\end{figure}
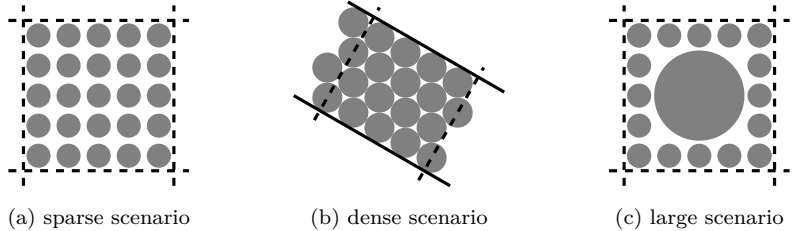

\begin{figure}
		\centering
		\subfloat[sparse scenario]{\includegraphics[width = 0.45\textwidth]{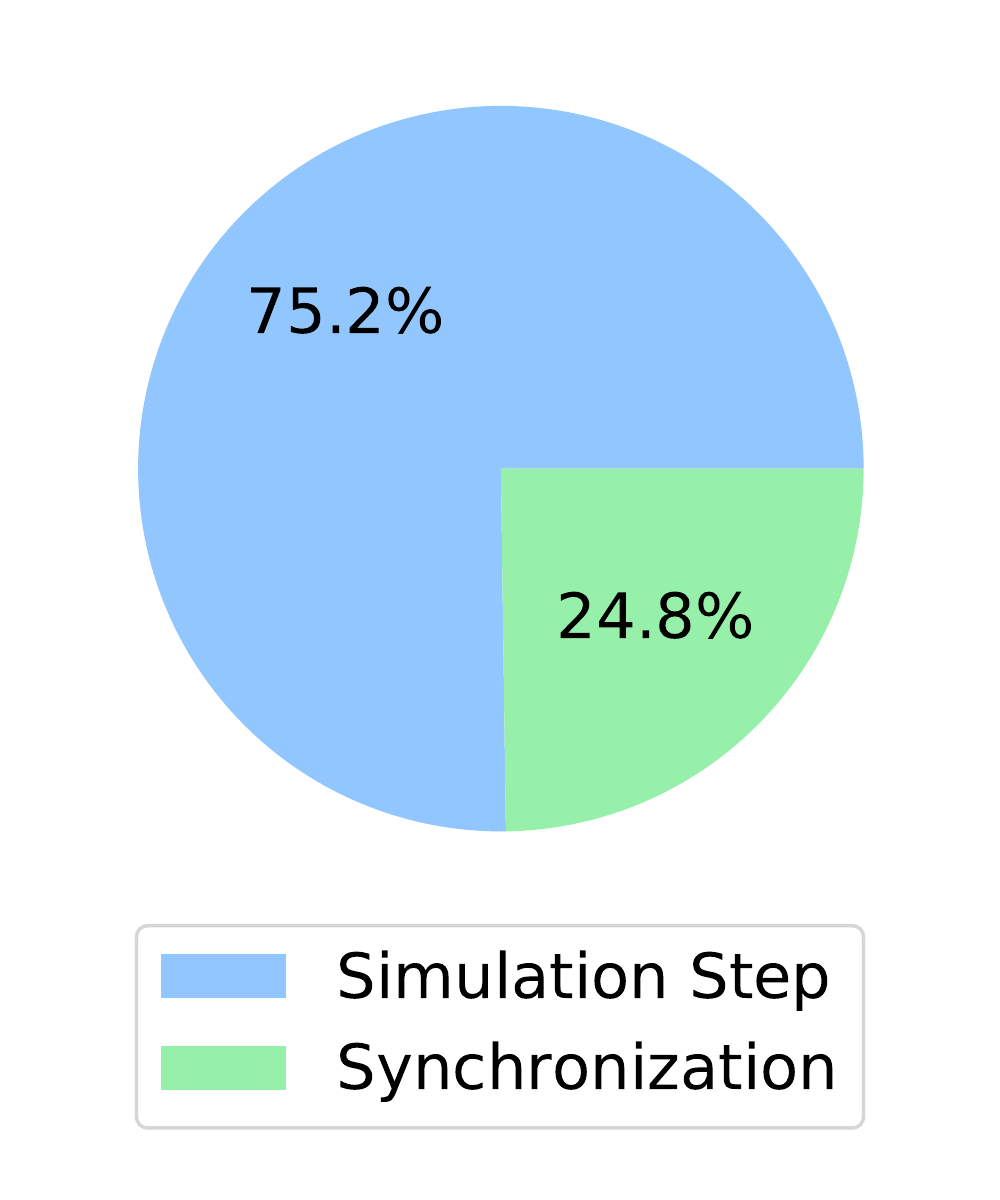}}
		\subfloat[dense scenario]{\includegraphics[width = 0.45\textwidth]{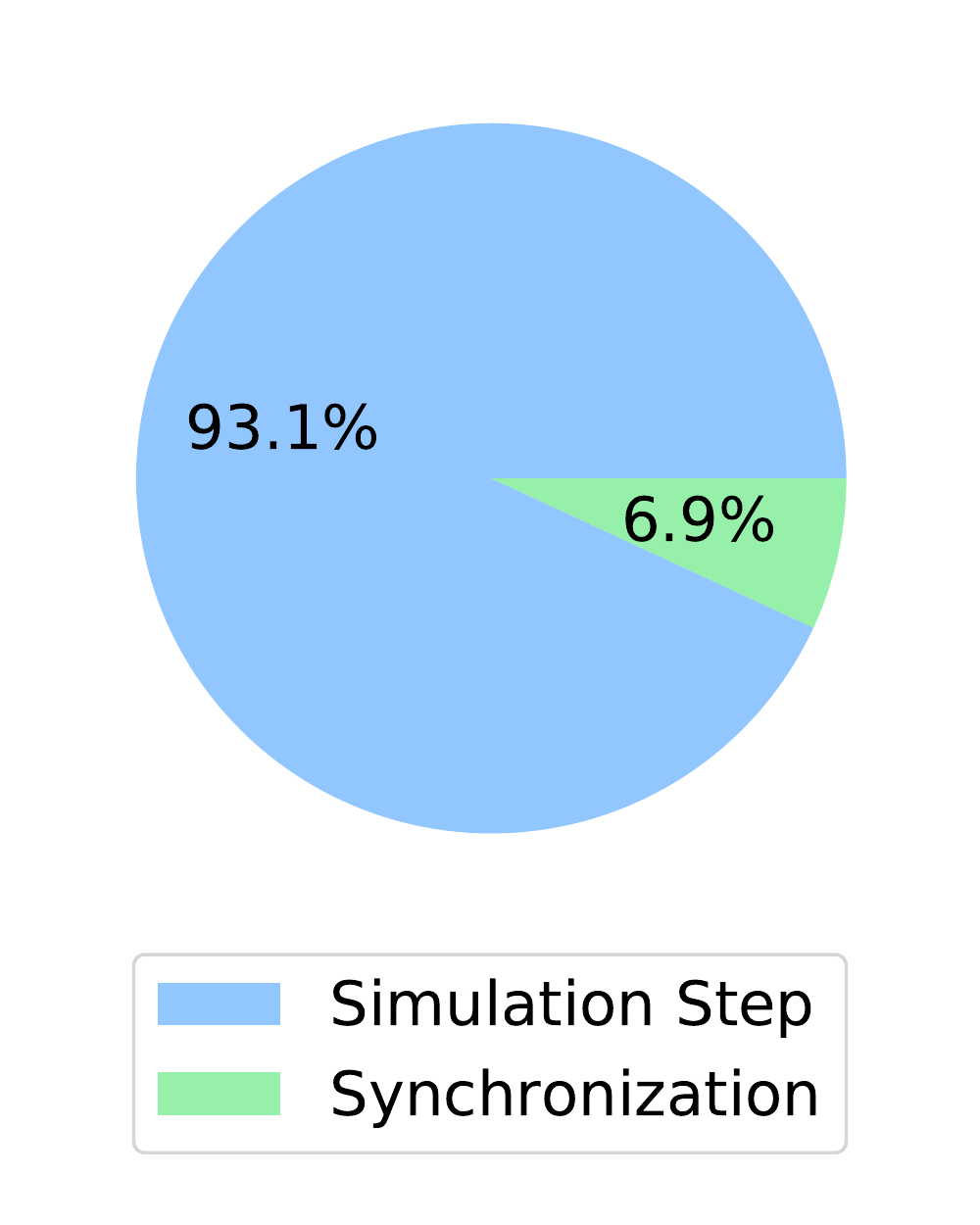}}	
		\caption{Depending on the relation between number of collisions and number of shadow particles, the synchronization can take up to one fourth of the total computation time. Both time measurements were taken from weak scaling runs using NNS with \num{1048576} processes.}
		\label{fig:RuntimeDistribution}
	\end{figure}

The simulation loop of a parallel rigid particle simulation can be divided into two phases: synchronization/communication and collision detection/resolution. The computation time spent in the collision detection/resolution phase heavily depends on the number of collisions which occur per time step. For a homogeneous scenario this corresponds to the particle density. To study the performance of the algorithms for a broad range of applications, we will investigate two different scenarios. One scenario is sparse with a low particle density where we expect that the synchronization and communication dominate the overall performance. In a second scenario we explore a dense setup where more compute time is spent in the collision detection and resolution. An exemplary distribution of runtime for two specific scenarios can be found in Fig.~\ref{fig:RuntimeDistribution}. An additional third scenario  with large particles is evaluated to demonstrate the features of our new synchronization algorithm. The values describing the scenarios in the upcoming paragraphs are given in dimensionless units. All scenarios kept simple to minimize the influence of other factors on our performance measurements.

\begin{figure}
		\centering
		\subfloat[sparse scenario, weak scaling]{\includegraphics[width = 0.45\textwidth]{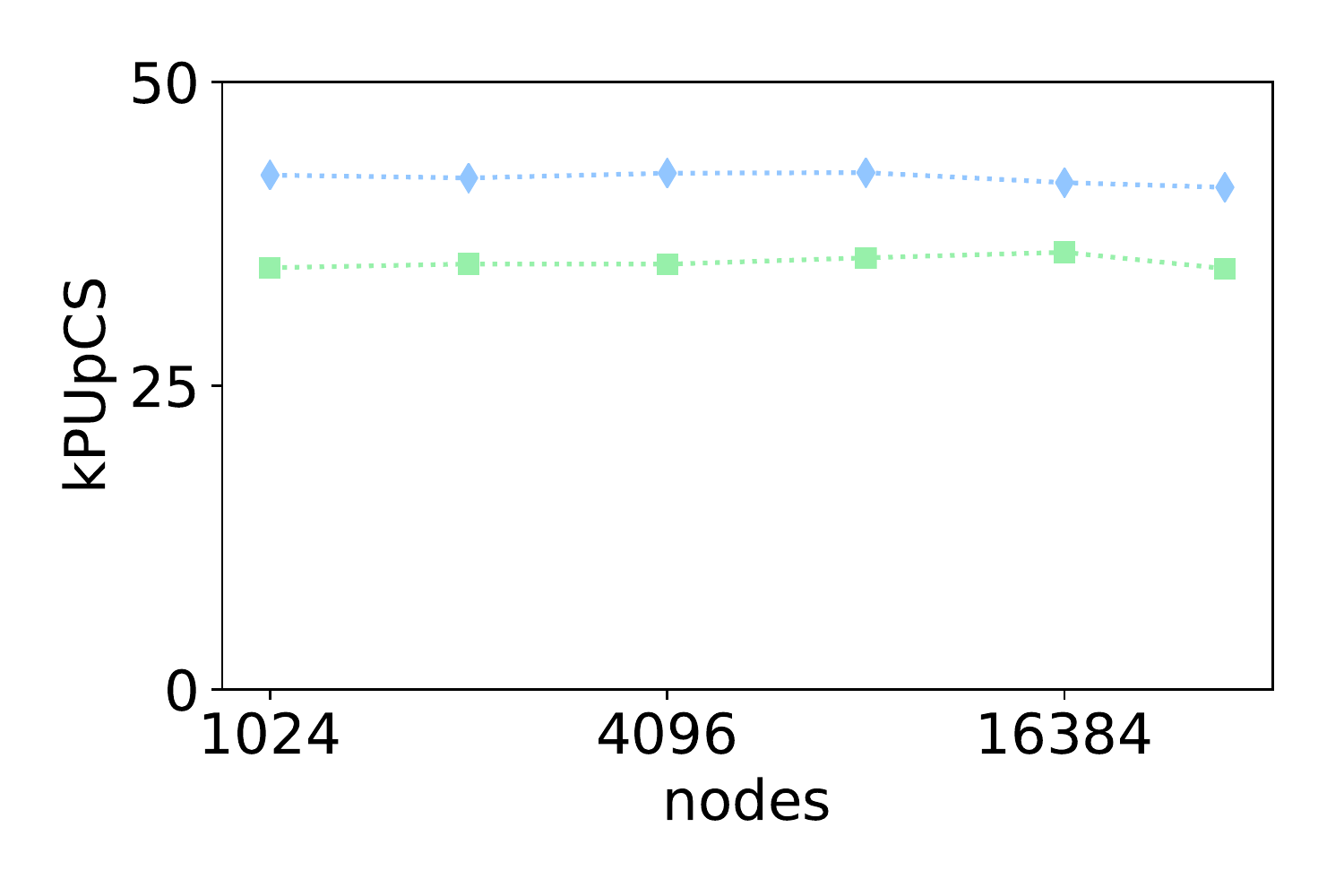}}
		\subfloat[sparse scenario, strong scaling]{\includegraphics[width = 0.45\textwidth]{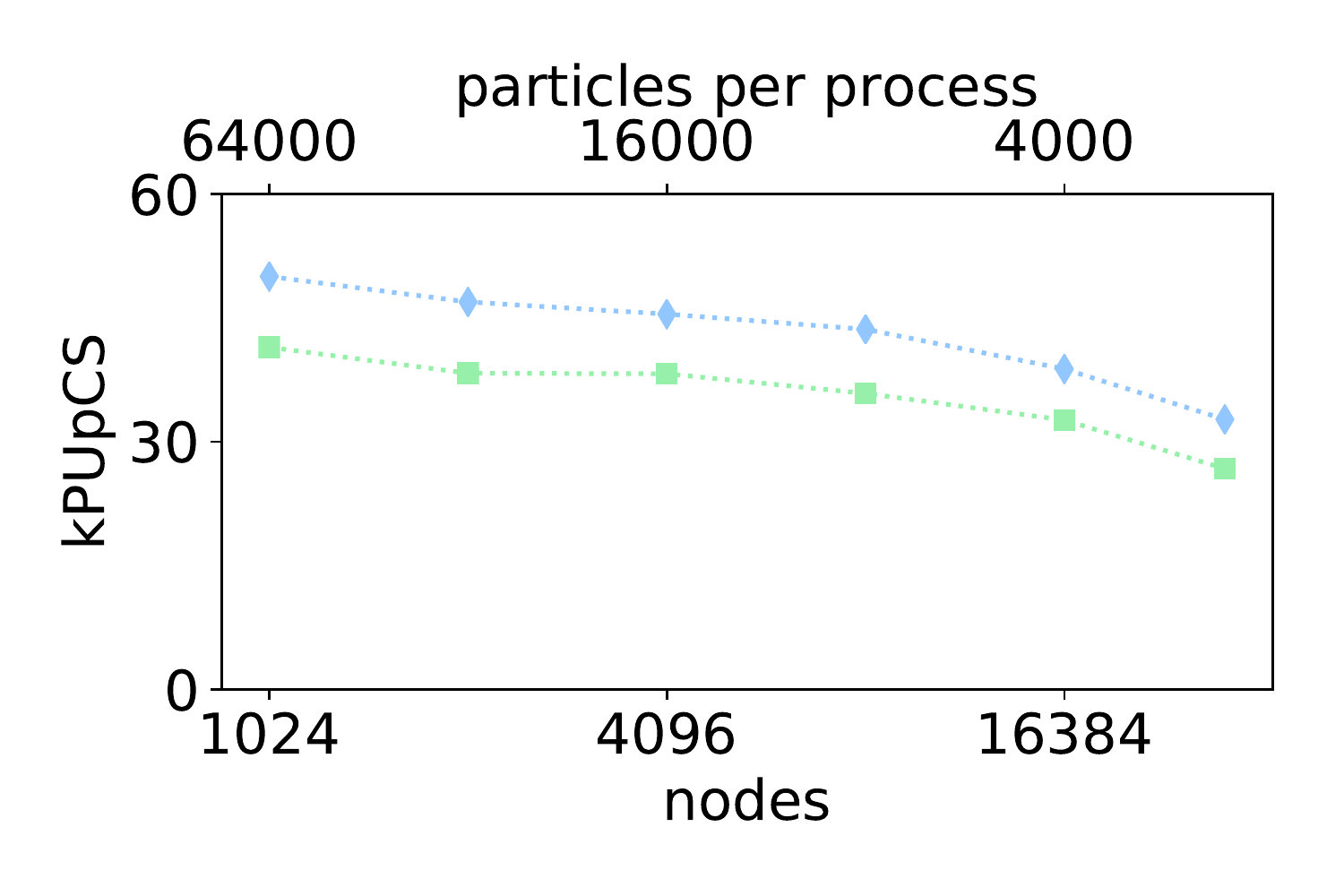}}

		\subfloat[dense scenario, weak scaling]{\includegraphics[width = 0.45\textwidth]{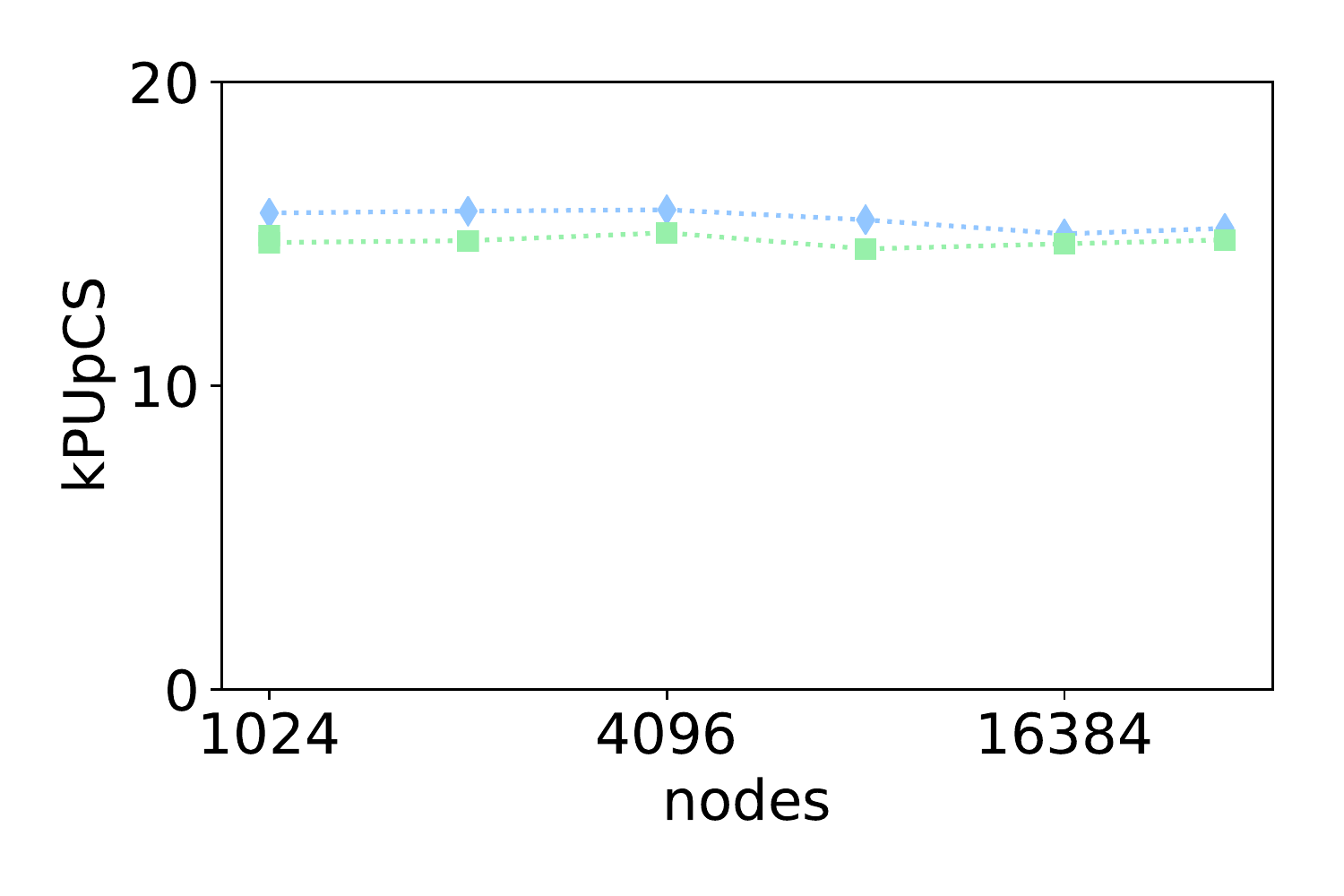}}
		\subfloat[dense scenario, strong scaling]{\includegraphics[width = 0.45\textwidth]{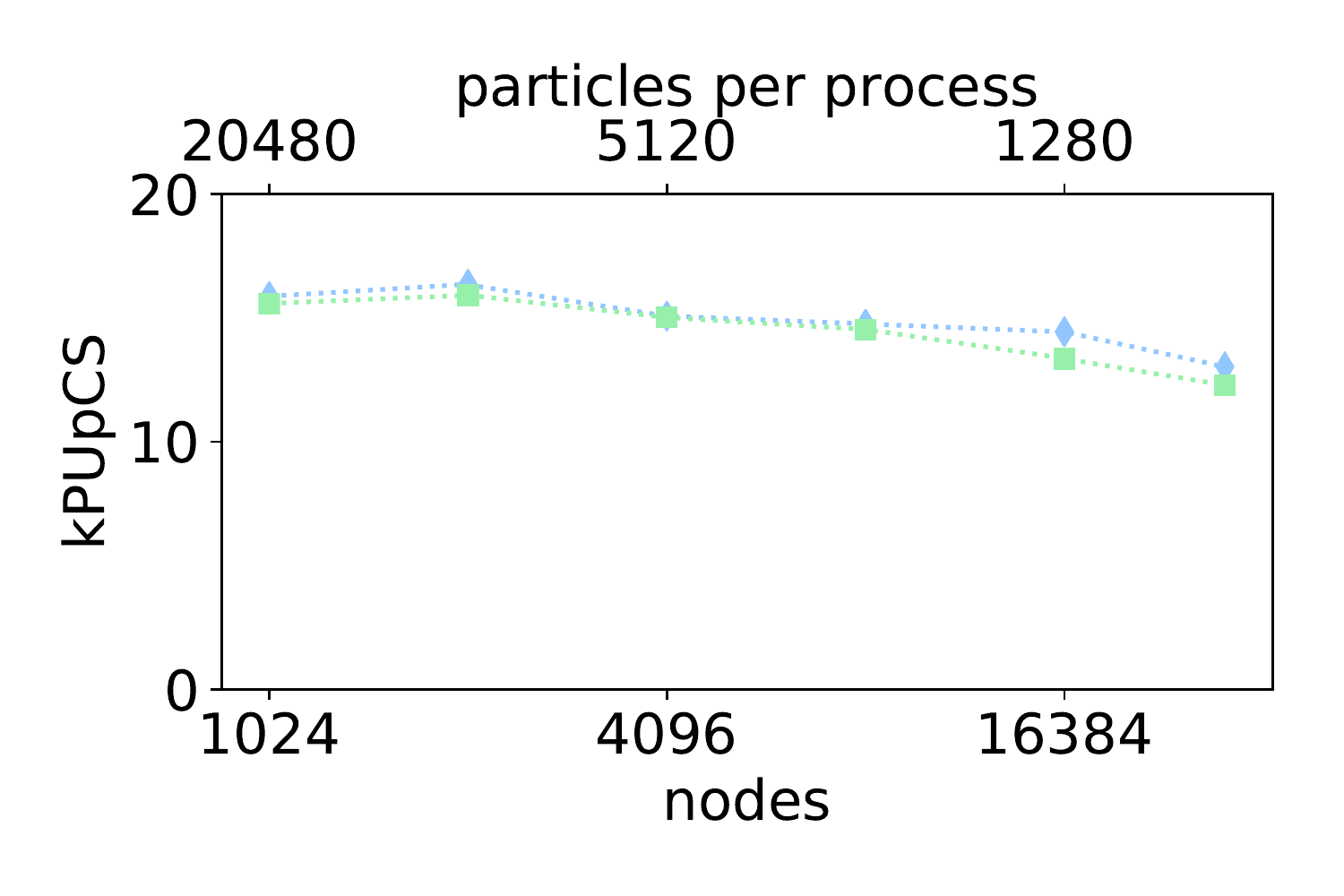}}

		\subfloat[large scenario, weak scaling]{\includegraphics[width = 0.45\textwidth]{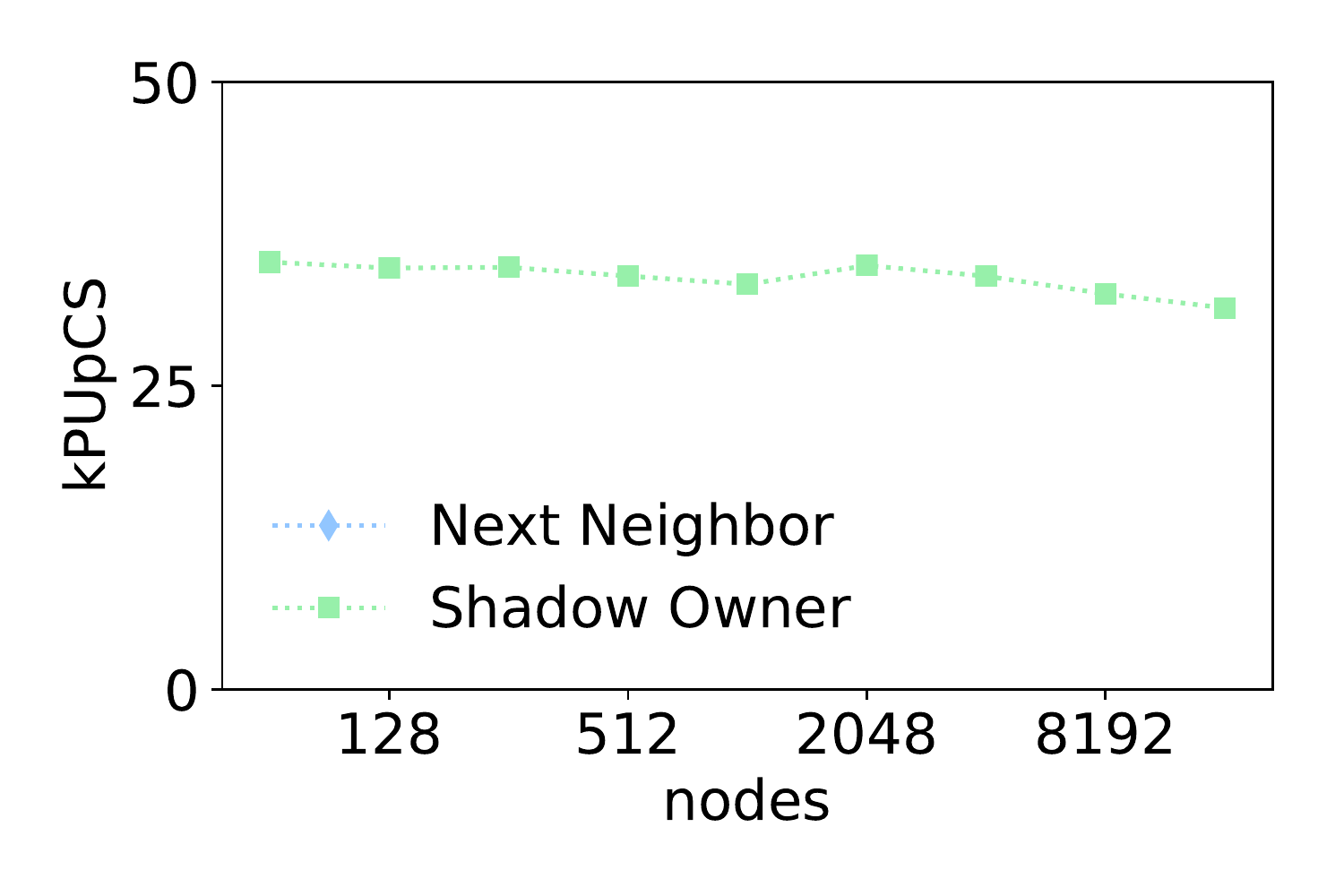}}
		\subfloat[large scenario, strong scaling]{\includegraphics[width = 0.45\textwidth]{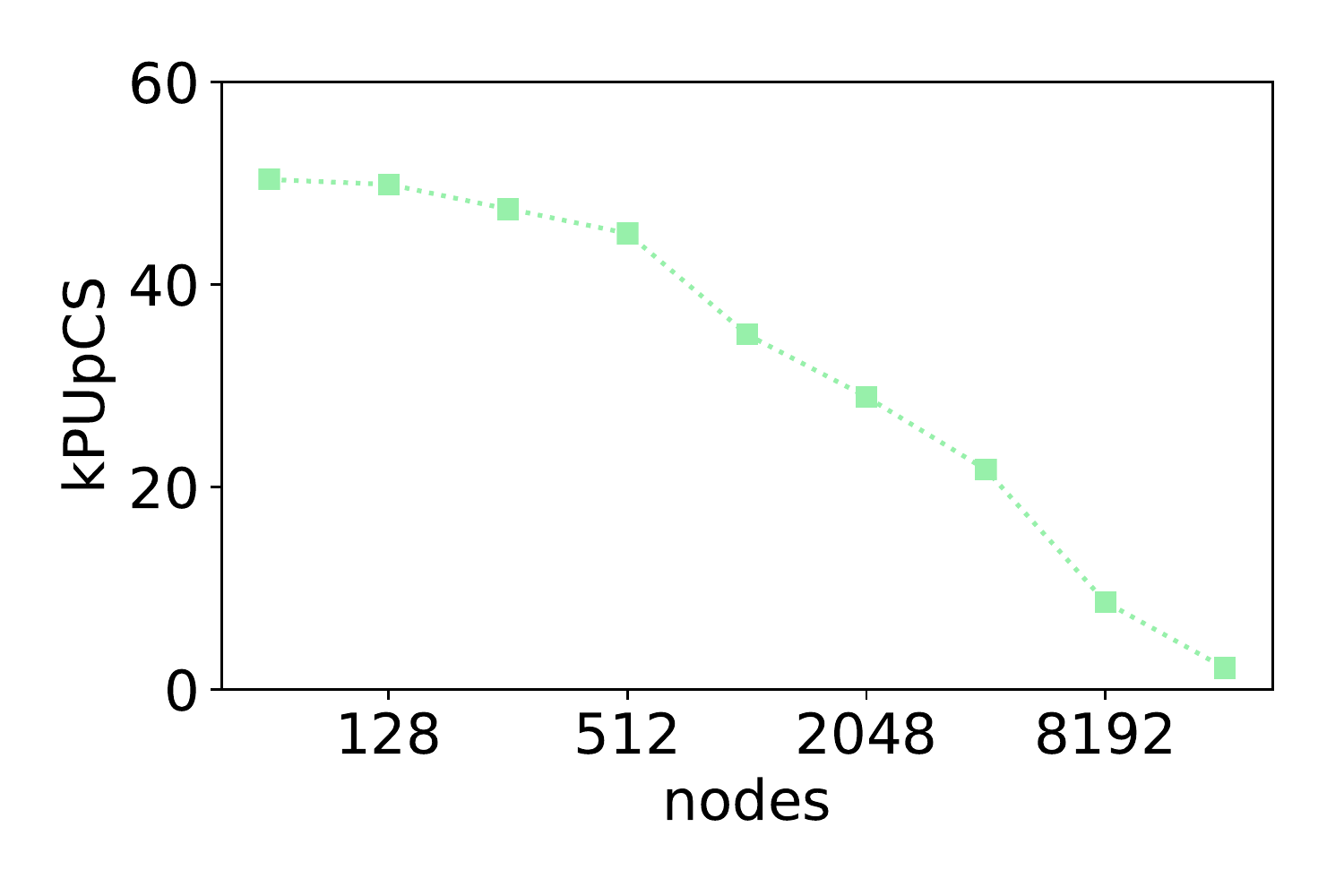}}

		\caption{In the left column the weak-scaling results are visualized for various scenarios. The right column show strong-scaling results. Each row corresponds to one scenario. The first row is a sparse granular gas scenario, the second row is a dense scenario and the last row is a bidisperse scenario which cannot be simulated with the traditional next neighbor approach. All scaling runs were conducted with both synchronization methods when possible. On the y-axis the achieved kilo particle updates per core second (see Sec.~\ref{sec:introduction}) are shown.}
		\label{fig:scaling}
	\end{figure}

\subsubsection{Sparse Scenario}
The sparse scenario is a granular gas confined in a static box. The granular gas consists of spherical particles. All particles are initialized on a rectangular grid (sc~\cite{J.R.Hook1991}) and have the same radius but a random initial velocity. The particle density is low enough for the particles to travel longer distances and to cross subdomain boundaries. Fig.~\ref{fig:Scenarios} a) shows a simplified illustration of this scenario. For the weak-scaling experiment a subdomain of size $20 \times 20 \times 20$ containing \num{8000} particles of radius \num{1} is assigned to every process. In our largest weak-scaling experiments we simulate \num{1.5e10} master particles and \num{1.8e9} shadow particles. This is executed on the full Juqueen with \num{28672} nodes using \num{1835008} processes on \num{458752} cores. For the strong scaling a fixed domain size of $\num{2560} \times \num{1280} \times \num{1280}$ is chosen. The domain is equally divided between all processes. This results in \num{64000} particles per process for the \num{1024} nodes run and \num{2285} particles for \num{28672} nodes run. More detailed parameters for the different scenarios can be found in Tab.~\ref{table:ScalingParameters}.

In this scenario both synchronization methods exclusively use next neighbor communication. Therefore we expect both algorithms to scale equally well. Fig.~\ref{fig:scaling} a) and b) show the results for the weak- and strong-scaling experiments using the sparse scenario. In the weak-scaling experiments neither of the algorithms drops in performance. Even on the whole machine with \num{1835008} processes both methods yield almost \SI{100}{\percent} parallel efficiency. The slow degradation in parallel efficiency for the strong-scaling gets noticeable above \num{8192} nodes (\num{524288} processes). During the increase from \num{1024} nodes to all available \num{28672} nodes the parallel efficiency drops to approximately \SI{70}{\percent} for both algorithms. This is due to the reduced number of particles per process. Fewer particles lead to fewer collisions and therefore proportionally less numerical effort is needed to advance the simulation. In this situation, algorithmic components that are independent of the number of particles like communication latency and logistic operations on the data structures become more prominent. These contributions dominate in strong-scaling experiments when the number of processes becomes very large. If we compare both algorithms we see that both methods exhibit the same behaviour. If we consider the absolute values of performance we can calculate that the more advanced SOS method achieves roughly \SI{83}{\percent} of the PUpCS of the NNS. Although the information communicated is the same for both synchronization methods, the SOS needs two communication cycles whereas the NNS needs only one. Therefore a slightly lower performance of the SOS as compared to the NNS is expected and is unavoidable.

\subsubsection{Dense Scenario}
The dense scenario consists of a periodic simulation domain confined in the z direction by solid walls. It is filled with spherical particles arranged on a hexagonal close packing (hcp~\cite{J.R.Hook1991}) lattice. The radius of the spheres is chosen such that the spheres touch their neighbors. Additionally a rotated gravitational force is applied such that the containing wall represents a ramp with \SI{30}{\degree} inclination. All spheres have an initial velocity down the ramp. This scenario is illustrated in Fig.~\ref{fig:Scenarios} b). This scenario is chosen because the distribution of computation time is different to the previous scenario. The communication is less significant since more of the computation time is spent inside the collision detection/resolution phase (see Fig.~\ref{fig:RuntimeDistribution}). For the experiment the simulation domain is partitioned only in the x and y direction since the height of the simulation stays the same. For the weak-scaling scenario each domain contains \num{4000} particles. The strong-scaling experiment uses the same scenario but with a simulation domain of fixed size. The number of particles per process ranges from \num{20480} particles on \num{1024} nodes to \num{731} on \num{28672} nodes.

Again we expect the same behaviour for both algorithms as only next neighbor communication is used. The graphs in Fig.~\ref{fig:scaling} c) and d) show the weak- and strong-scaling. The weak-scaling for both algorithms shows no performance loss even for the full-machine runs. For the strong-scaling case the parallel efficiency drops to approximately \SI{78}{\percent} for both algorithms in the largest simulations. Compared to the sparse scenario this is significantly less. Since the computation time of the simulation is directly related to the number of collisions, each particle requires more computation time in the dense scenario. Therefore particle independent effects like latency and maintenance start to dominate at fewer particles leading to a better overall strong scaling. Looking at the absolute performance difference between the two algorithms we can conclude that for dense particle ensembles the improved functionality of SOS induces only insignificant extra costs.

\subsubsection{Large Particle Scenario}
Since the SOS was specifically developed to tackle problems with large particles one scenario of this category is also evaluated. NNS cannot be applied here therefore only the SOS is studied. For a bidisperse scenario with spherical particles of two radii is chosen. The smaller particle always has radius 1 whereas the larger particle has a radius of 30 in the weak-scaling experiment and 15 in the strong-scaling experiment. A simulation area of $80 \times 80 \times 80$ is used as a building block. The complete simulation domain is then constructed by placing these building blocks adjacent to each other. The building block is depicted in Fig.~\ref{fig:Scenarios} c). It is partitioned into $4 \times 4 \times 4$ subdomains resulting in subdomains with an edge length of \num{20}. Each subdomain is assigned to exactly one process. In the center of this building block a large spherical particle is introduced. The rest of the simulation area is filled with small particles arranged on a simple cubic lattice. All particles are given a random initial velocity. 

For this scenario the results for weak- and strong-scaling can be found in Fig.~\ref{fig:scaling} e) and f). The SOS shows superb weak-scaling also in this kind of scenario up to the full machine. In the strong-scaling case one can observe a breakdown in performance starting at \num{512} nodes. This can be explained as follows. Since the large particle is positioned at a 3d corner where eight subdomains meet it overlaps all of them. For \num{512} nodes each subdomain is $20 \times 20 \times 20$ in size. From now on every further refinement of the domain leads to an increase of the number of subdomains that the particle overlaps. As a consequence the communication volume increases considerably and the performance deteriorates.

Note that the absolute PUpCS depend on the number of collisions per particle. Since the time needed for collision resolution is directly related to the number of collisions a dense scenario will always achieve less PUpCS than a sparse scenario. This can also be seen in Fig.~\ref{fig:scaling}.

	\begin{sidewaystable}
		\centering
		\begin{tabular}{llll}
			\toprule
			parameter & sparse & dense & large \\
			\midrule
			weak-scaling domain size (per process) & $20 \times 20 \times 20$ & $20 \times 17.3 \times 8$ & $20 \times 20 \times 20$ \\
			strong-scaling domain size (total) & $2560 \times 1280 \times 1280$ & $4096 \times 3547 \times 8$ & $640 \times 640 \times 640$ \\
			domain partitioning & 3d & 2d & 3d \\
			lattice structure~\cite{J.R.Hook1991} & sc & hcp & sc \\
			grid spacing & 1 & 1 & 1 \\
			sphere radius & 0.4 & 0.5 & 0.4 \\
			large particle radius & - & - & 30(15) \\
			initial $v_\text{max}$ & $\left[-0.1, +0.1\right]$ & 0.1 & $\left[-0.1, +0.1\right]$ \\
			time step length & 1 & 0.01 & 0.1 \\
			time steps & 1000(500) & 1000(300) & 1000(100) \\
			\bottomrule
		\end{tabular}
		\caption{Simulation parameters for each scenario. Values given in brackets refer to the strong scaling case. If a range is given for $v_\text{max}$ it means that every component of the vector is a random value in that range. The number of time steps refers to the number of time steps used to build the average of the simulation time used.}
		\label{table:ScalingParameters}
	\end{sidewaystable} 

\subsection{Direct Comparison of both Algorithms}
	\label{sec:Comparison}
	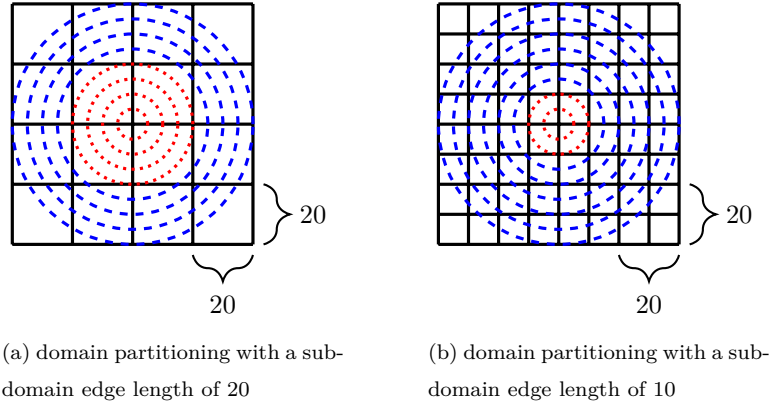
\begin{figure}
		\centering

		\subfloat[domain partitioning with a subdomain edge length of 20]{
		\begin{tikzpicture}[scale=0.04]
			\draw [black, step=20.0, very thick] (0.0,0.0) grid (80,80);
			\draw [red, dotted, very thick] (40,40) circle [radius=20];
			\draw [red, dotted, very thick] (40,40) circle [radius=15];
			\draw [red, dotted, very thick] (40,40) circle [radius=10];
			\draw [red, dotted, very thick] (40,40) circle [radius= 5];
			\draw [black, thick, decorate, decoration={brace,amplitude=10pt,mirror,raise=4pt},yshift=0pt] (80, 0) -- (80,20) node [black,midway,xshift=+0.8cm] {20};
			\draw [black, thick, decorate, decoration={brace,amplitude=10pt,mirror,raise=4pt},yshift=0pt] (60, 0) -- (80, 0) node [black,midway,yshift=-0.8cm] {20};

			\draw [blue, dashed, very thick] (40,40) circle [radius=40];
			\draw [blue, dashed, very thick] (40,40) circle [radius=35];
			\draw [blue, dashed, very thick] (40,40) circle [radius=30];
			\draw [blue, dashed, very thick] (40,40) circle [radius=25];

		\end{tikzpicture}
		}
		\hspace{1cm}
		\subfloat[domain partitioning with a subdomain edge length of 10]{
		\begin{tikzpicture}[scale=0.04]
			\draw [black, step=10.0, very thick] (0.0,0.0) grid (80,80);
			\draw [red, dotted, very thick] (40,40) circle [radius=10];
			\draw [red, dotted, very thick] (40,40) circle [radius= 5];
			\draw [black, thick, decorate, decoration={brace,amplitude=10pt,mirror,raise=4pt},yshift=0pt] (80, 0) -- (80,20) node [black,midway,xshift=+0.8cm] {20};
			\draw [black, thick, decorate, decoration={brace,amplitude=10pt,mirror,raise=4pt},yshift=0pt] (60, 0) -- (80, 0) node [black,midway,yshift=-0.8cm] {20};

			\draw [blue, dashed, very thick] (40,40) circle [radius=40];
			\draw [blue, dashed, very thick] (40,40) circle [radius=35];
			\draw [blue, dashed, very thick] (40,40) circle [radius=30];
			\draw [blue, dashed, very thick] (40,40) circle [radius=25];
			\draw [blue, dashed, very thick] (40,40) circle [radius=20];
			\draw [blue, dashed, very thick] (40,40) circle [radius=15];

		\end{tikzpicture}
		}

		\caption{Illustration of the domain partitioning used for the bidisperse setup. The extension of the large centered sphere for different configurations is indicated by additional circles. All configurations (dashed and dotted circles) can be simulated with the proposed SOS whereas only the dotted red ones can be simulated with the traditional NNS.}
		\label{fig:SimulationArea}
	\end{figure}

\begin{figure}
		\centering
		\includegraphics[width = 0.8\textwidth]{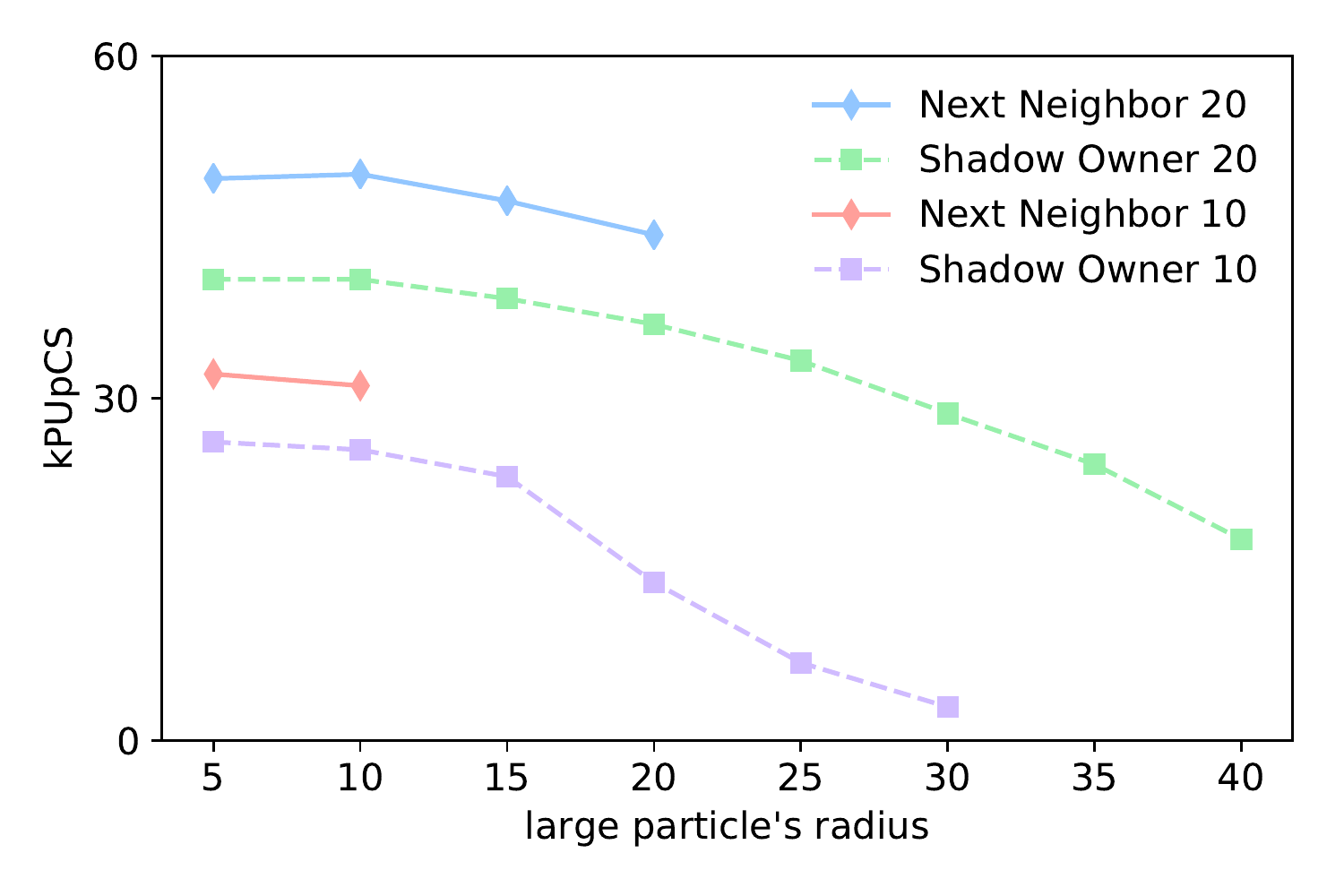}
		\caption{Both synchronization methods are compared in a bidisperse scenario for two different domain partitionings. The edge length of one subdomain is given behind the name of the synchronization method. The radius of the large particles is varied and given on the x axis. The surrounding area is sparsely filled with spherical particles of radius \num{0.4}.}
		\label{fig:SyncProtocolSparse}
	\end{figure}

In this section we evaluate the parallel performance for different particle size ratios. This demonstrates the extended functionality of SOS and at the same time allows to examine the performance cost in more detail. The setup is similar to the third scenario of the last paragraph. The same building block is used but the radius of the large particle in the center is varied. To reduce fluctuation of the time measurement a grid of $4 \times 4 \times 4$ of these building blocks is simulated. We use the same domain partitioning into $4 \times 4 \times 4$ subdomains per building block as in the third scenario of the last paragraph and add an additional experiment with a finer partitioning of $8 \times 8 \times 8$. The domain partitioning as well as the different sizes of the large particle are illustrated in Fig.~\ref{fig:SimulationArea}. All particle sizes can be simulated with SOS but only those large particles which are marked by a dotted red line can be simulated with NNS. As long as the radius of the large sphere is not larger than the edge length of one subdomain both synchronization methods can be used. When the radius reaches a certain value only the SOS can be used. These radii are indicated by blue dashed circles. For the first domain partitioning into $4 \times 4 \times 4$ subdomains \num{4096} processes are needed to run and for the second one \num{32768}. All possible setups are run and the achieved performance is compared in Fig.~\ref{fig:SyncProtocolSparse}. 

As the previous experiments already suggest, the SOS achieves about \SI{80}{\percent} of the PUpCS of NNS. However SOS can be applied to a wider range of scenarios. This can be seen by the fact that for SOS all data points are available whereas the graphs for NNS stop when it is impossible to simulate the next particle radius. The overall performance decreases moderately when the radius of the large particle increases. A large particle radius leads to an increased amount of communication as it overlaps more subdomains. Note also that processes whose subdomains are completely covered by a large particle will be idle which further decreases the performance. This effect could be overcome by sophisticated load balancing algorithms which however are outside the scope of this paper. The performance also drops if we progress to a finer domain subdivision. For a subdomain edge length of \num{20} roughly \num{8000} particles are located in each subdomain. If we use a partitioning with a subdomain edge length of \num{10} only \num{1000} particles remain per subdomain. When the workload per process is so small, a further loss in efficiency is unavoidable (compare Fig.~\ref{fig:scaling}b)).

	\begin{figure}
		\centering
		\includegraphics[width = 0.8\textwidth]{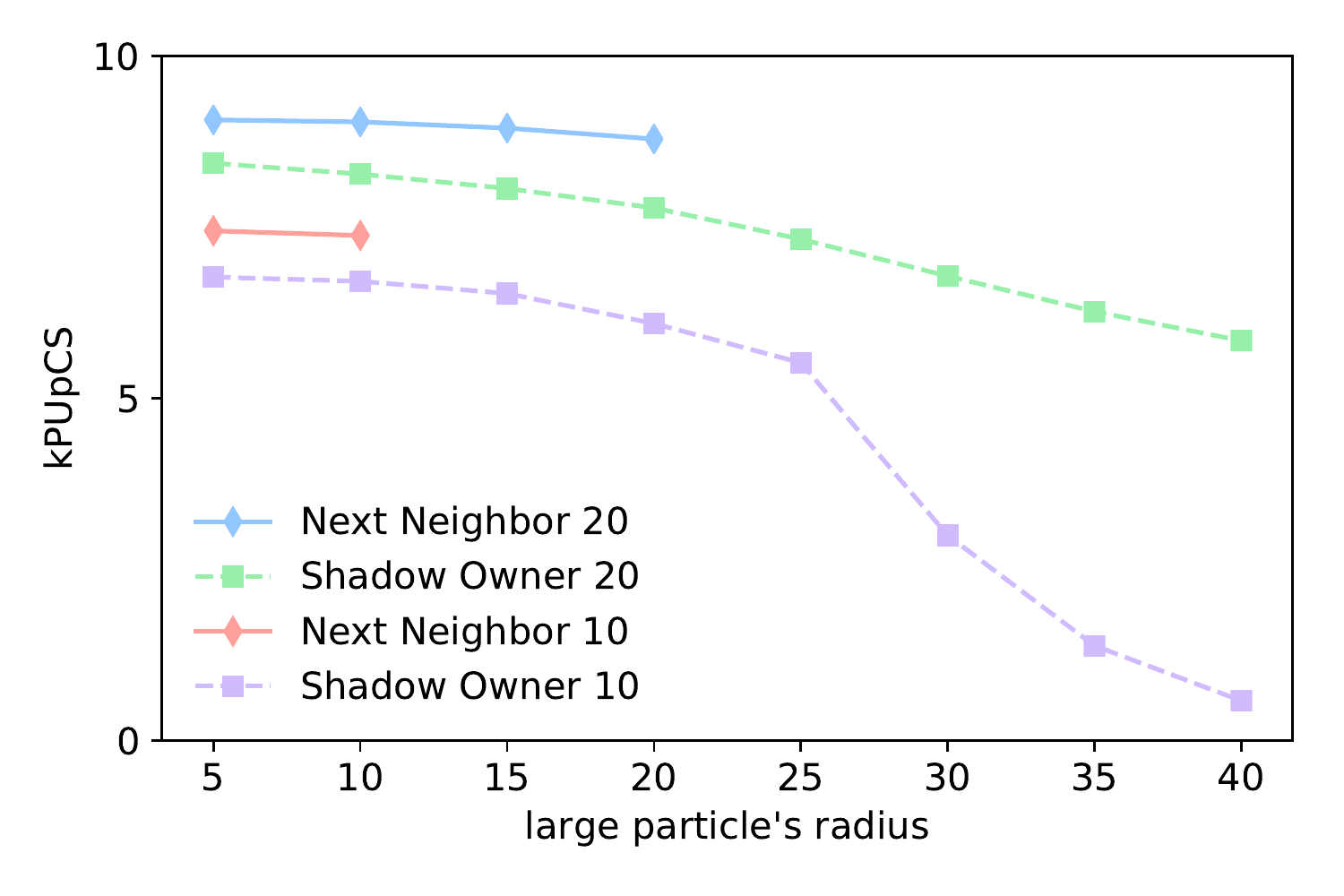}
		\caption{The same setup is used as in Fig.~\ref{fig:SyncProtocolSparse} but the area surrounding the large particles is now densely filled with spherical particles of radius \num{0.98}.}
		\label{fig:SyncProtocolDense}
	\end{figure}

We repeat the measurements with a denser setup. For that the small spherical particles are arranged on a hexagonal close packing lattice with spacing \num{1} and a radius of \num{0.4}. This setup consists of slightly more particles and also the number of collisions is higher due to the dense packing. Fig.~\ref{fig:SyncProtocolDense} shows the performance data for this setup. As anticipated form the previous scaling experiments both synchronization methods yield almost the same PUpCS. The SOS achieves roughly \SI{90}{\percent} of the PUpCS of NNS. And also the general trend is the same as for the sparse setup due to the same reasons.

\section{Conclusion}
\label{sec:Conclusion}
In this paper we have presented a solution to resolve a limitation of the NNS algorithm used in rigid body dynamics. The NNS only exchanges information with next neighbors. Therefore no particle is allowed to extend into subdomains past direct adjacent ones. We have proposed a new synchronization algorithm which uses a minimal amount of point-to-point communications to exchange the needed information even with subdomains which are not adjacent. New shadow particles are created with a diffusive approach. Careful bookkeeping and caching are applied to efficiently implement this algorithm. We then have shown that this algorithm scales as well as the NNS in various scenarios up to \num{1835008} processes on the Juqueen supercomputer. We have also shown that for dense scenarios the NNS method can be replaced by our newly proposed SOS with only a moderate performance penalty. Eventually we have applied our new method to scenarios which previously could not have been simulated and showed a broader range of simulations which are now possible. Further improvements could be gained by introducing advanced load balancing techniques reducing the number of of idling processes in scenarios with large particles. Since the synchronization protocol is independent of the contact model both the NNS as well as our new SOS can be used in conjunction with various interaction solvers. This enables our synchronization method to be easily adapted to for example molecular dynamics. 

\section*{Acknowledgment}
The authors gratefully acknowledge the Gauss Centre for Supercomputing e.V. (www.gauss-centre.eu) for funding this project by providing computing time through the John von Neumann Institute for Computing (NIC) on the GCS Supercomputer JUQUEEN at Jülich Supercomputing Centre (JSC).

The authors would like to acknowledge the support through the Cluster of Excellence Engineering of Advanced Materials (EAM).

Declaration of interest: none

\section*{References}

\bibliography{communication}

\end{document}